\documentclass[12pt,english]{svjour3}
\usepackage{mathptmx}

\usepackage[T1]{fontenc}
\usepackage[latin9]{inputenc}
\usepackage[letterpaper]{geometry}
\usepackage{color}
\usepackage{prettyref}
\usepackage{float}
\usepackage{calc}
\usepackage{url}
\usepackage{amsmath}
\usepackage{amssymb}
\usepackage{graphicx}
\usepackage{esint}

\makeatletter
\RequirePackage{fix-cm}

\smartqed  % flush right qed marks, e.g. at end of proof

\usepackage{textcomp}

\usepackage[justification=centering]{caption}

\usepackage{float}
\makeatother

\usepackage{babel}
\begin{document}

\title{{\large{}On the Mössbauer effect and the rigid recoil question}}

\subtitle{preprint version 1}

\author{\textup{\normalsize{}Mark Davidson}}

\institute{Spectel Research Corporation; Email: mdavid@spectelresearch.com}

\date{{\normalsize{}June 17, 2016}}
\maketitle
\begin{abstract}
Various theories for the Mössbauer rigid-recoil effect, which enables
a crystal to absorb momentum but not appreciable energy, are compared.
These suggest that the recoil may not be instantaneous, and that the
recoil time could be used to distinguish between them. An experiment
is proposed to measure this time. The idea is to use a small sphere
whose outer surface is coated with an electrically charged Mössbauer-active
element, and then to measure the amount of energy lost due to Bremmsstrahlung
during the recoil of this sphere when a Mössbauer event occurs. As
the energy radiated is proportional to the square of the acceleration
from Larmor's formula, the amount of energy so radiated varies inversely
proportional to the recoil time, and proportional to the charge squared.
Although this energy is quite small, it can in principle be measured
with the extreme sensitivity available in Mössbauer experiments. It
is found that the most information would be gained with more long-lived
isomers such as Rhodium-103m as the Mössbauer agent, and that correction
terms to Larmor's formula need to be considered due to the finite
radius of a small particle, and some of these corrections are included
in the paper. These corrections are taken from the extensive literature
on radiating spherical shells which were developed to analyze the
Abraham-Lorentz model for a classical charged particle. This problem
is interesting because a perfectly rigid crystal seems to pose questions
about relativity, and because the Mössbauer effect is a phenomenon
where the solid-state environment dramatically influences the probability
of a nuclear process. Moreover, if the experiment proposed can measure
the recoil time, then such measurements might provide a new class
of analytical methods for material science and chemistry.
\end{abstract}

\section{{\normalsize{}Introduction}}

In the Mössbauer effect, a nucleus emits or absorbs gamma rays without
the expected loss of energy due to the recoil of the decaying or the
absorbing nucleus \cite{mossbauer_kernresonanzfluoreszenz_1958,frauenfelder_mossbauer_1962,lipkin_quantum_2014}.
It is extraordinarily sensitive and has found many uses in chemistry,
condensed matter physics, and tests of relativity \cite{chen_mossbauer_2007,frauenfelder_mossbauer_1962,fultz_mossbauer_2011,gibb_principles_2013,greenwood_mossbauer_1971,long_mossbauer_2013,morup_mossbauer_1990,sharma_mossbauer_2013}.
The conventional explanation for this effect is that in the right
circumstances a volume of a crystal in which the decaying nucleus
is embedded can act effectively like a rigid body in response the
nuclear decay, and since the crystal's mass is much larger than that
of one nucleus alone, it can absorb the recoil momentum of a gamma
ray without absorbing any appreciable amount of energy \cite{frauenfelder_mossbauer_1962,lipkin_quantum_2014,bressani_what_1992}.
The same can happen on both emission and absorption. The probability
of recoilless decay and resonant absorption depends on the temperature,
on the phonon density of states, and on the detailed crystal structure
\cite{frauenfelder_mossbauer_1962,chen_mossbauer_2007,lipkin_quantum_2014,lipkin_simple_1960}.
In classical relativity theory, it is generally conceded that a perfectly
rigid body cannot exist because it would allow superluminal messages
to be sent from one end of the body to another. There is some acknowledgement
in the theoretical literature that the recoil may not, and perhaps
cannot be instantaneous throughout a volume of the crystal. For example,
Victor Weisskopf argued that the diffusion velocity of energy in a
solid is limited by the sound velocity in that solid, and that consequently
the effective radius of a rigid crystal would be given by the speed
of sound times the lifetime of the excited isomer undergoing decay
\cite{weisskopf_selected_1961}. Fruaenfelder acquiesces to this viewpoint
in his theoretical discussion in chapter 2 of \cite{frauenfelder_mossbauer_1962}.
Another viewpoint comes from particle physics. In the standard model,
the energy-momentum density $T^{\mu\nu}(x)$ is locally conserved
in any Lorentz covariant classical field theory. Thus the maximum
propagation is taken to be the speed of light, and there is therefore
no reason why energy transport cannot occur up to this speed in a
solid. Quantum electrodynamics is considered to be a local theory
because the commutator of any two Bosonic fields separated by spacelike
separation vanishes, although the precise statement of locality is
somewhat problematical in this case since the Feynman propogators
(for the photon for example) do not vanish outside of the light cone.
The application of the standard model to a solid is not easy, and
this issue of locality is not completely settled. In general, I would
venture a guess that most theoretical physicists would believe the
rigid recoil picture to be incompatible with the perturbative picture
of local quantum field theory and thus with the standard model, but
non-perturbative rigid behaviour might be possible, as was argued
by Preparata in his superradiance theory \cite{bressani_what_1992,preparata_qed_1995}.
The standard model, which includes QED, is believed to be the best
fundamental theory of all matter at this time. So we have three possibilities
in the theoretical literature
\begin{enumerate}
\item The recoil momentum is absorbed locally, and the added momentum diffuses
throughout the crystal at the speed of sound
\item The recoil momentum is absorbed locally, and the momentum diffuses
at the speed of light
\item The recoil momentum is absorbed non-locally and the whole crystal
recoils instantaneously
\end{enumerate}
So which case does nature choose? Or does this question even make
any sense? Can an experiment be devised to answer it?

When a Mössbauer crystal recoils rigidly, it accelerates for a short
time. In the rigid body picture, the acceleration is very large, but
in a local field theory model the expected acceleration is much smaller,
as the momentum injected into the crystal at the location of the decaying
nucleus would need time to diffuse throughout the rest of the solid.
If the crystal has a net charge, then it will radiate electromagnetic
radiation due to this acceleration. If the crystal is metallic, then
the charge will reside on the outside surface. To a first approximation,
Larmor's radiation formula can be considered. This leads to a different
radiation loss between instantaneous acceleration models and slower
acceleration models, because the radiated power is proportional to
the acceleration squared. Standard Mössbauer spectroscopy can be used
then to try and measure the radiation loss with exquisite precision,
and thus to determine which model for acceleration fits the data best. 

The purpose of this paper is to examine this apparent contradiction
between the three viewpoints regarding locality and diffusion rates
after Mössbauer events, and to propose an experiment that might help
to clarify the issue, and which might also provide new laboratory
methods of practical use. We consider a metallic nano-particle with
an electromagnetic charge on its surface as either the source or the
absorber in a Mössbauer experiment. When the nucleus either emits
or absorbs a gamma ray, the crystal will recoil. It will experience
acceleration for a short time, and there will be energy loss due to
low frequency electromagnetic radiation. Using Larmor's formula to
estimate this radiation, we show that the amount of energy radiated
depends on the charge squared, and also on how abruptly the crystal
recoils. A perfectly rigid crystal which accelerated instantaneously
would have a very large energy loss, and a slow diffusive recoil mediated
by sound waves would have a much lower loss. If the diffusion occurred
at near to the speed of light, the loss would be somewhere in between.
The energy loss can be measured in the Mössbauer setup, and thus an
experimental measurement can in principal be made of the recoil dynamics. 

The Mössbauer effect is interesting for two fundamental reasons, aside
from its many practical uses. First, it raises these fundamental question
about locality, and second because it illustrates a system in which
the chemical environment indisputably has a very large effect on the
reaction rate of a nuclear process.

\section{{\normalsize{}Textbook theories of the Mössbauer effect from condensed
matter physics}}

There are a number of theoretical approaches to describing the Mössbauer
effect in the literature. Many historical papers can be found in Hans
Frauenfelder's excellent book \cite{frauenfelder_mossbauer_1962}.
The basic physical picture is that a region of a crystal recoils as
a rigid body in response to a nuclear decay. The solid is typically
approximated by the Debye model which treats it as a set of coupled
oscillators. Quantization of these oscillators leads to a phonon spectrum
with a non-zero energy gap between the ground state and the next excited
state. When the recoil energy of the Mössbauer is comparable to this
energy gap, then the probability for recoilless decay can become appreciable,
depending on the Debye-Waller factor. Frauenfelder acknowledges the
possible existence of a relaxation process of non-zero time duration
after the nuclear decay but before the final state of the rigidly
moving crystal is realized, but he does not go into details (see chapter
2 of \cite{frauenfelder_mossbauer_1962}) of this transient possibility.
His discussion of transient effects is influenced by a paper by Weisskopf
\cite{weisskopf_selected_1961}. Many papers simply treat the recoil
as if it were an instaneous rigid recoil, and they judiciously avoid
discussing any transient response of the system. The subject of rigid
bodies and special relativity is discussed in some detail in \cite{franklin_rigid_2013},
and the possibility of a rigid recoil of the whole crystal does not
seem to be automatically ruled out. All of the early theories for
the Mössbauer effect were developed before the standard model was
known. An attempt was made to reconcile instantaneous recoil with
the standard model in \cite{bressani_what_1992,preparata_qed_1995}
where the authors argued that the possibility of superradiant coherent
oscillation of a group of nuclei at the plasma frequency might explain
the rigidity of the crystal. I want to concentrate here on the rigid-recoil
aspect of these theories, and so a simplified treatment is in order.
The treatment by Lipkin \cite{lipkin_simple_1960,lipkin_quantum_2007}
is a good starting point to examine the locality of this process without
getting into unnecessary complications. First of all, as in almost
all theoretical models in solid state physics, the treatment is always
done with non-relativistic quantum mechanics, the argument being that
since the velocities of the nuclei remain non-relativistic after the
decay, this is justified. This allows us to have a nice many-body
Schrödinger picture with a local Hamiltonian description and a well
defined Hilbert space of states. Most quantum mechanical discussions
of the Mössbauer effect start by considering an isolated and free
nucleus experiencing decay. The discussion typically proceeds something
like this. Let $\left|in\right\rangle $ and $\left|out\right\rangle $
denote the nuclei's wave function before and after the decay emission
of a gamma ray, and we know that both that the total kinetic energy
and momentum are conserved in the decay process. Furthermore, let
us assume that the initial state is an eigenstate of momentum $\left|in\right\rangle =\left|\mathbf{p_{in}}\right\rangle $.
If we denote the momentum of the decay gamma ray as $\mathbf{p}_{\gamma}$,
then we have $\left|out\right\rangle =\left|\mathbf{p}_{in}-\mathbf{p}_{\gamma}\right\rangle $.
The rest mass of the decaying nucleus must change to the mass of the
nuclear ground state that it decays into. Call these masses $M_{in}$
and $M_{out}$. These are known fixed values which can be found in
standard nuclear databases. The kinetic energy of the nucleus before
decay is $E_{in}=\mathbf{p}_{in}^{2}/2M_{in}$, and the kinetic energy
after the decay is $E_{out}=\left(\mathbf{p}_{in}-\mathbf{p}_{\gamma}\right)^{2}/2M_{out}$.
Conservation of total energy requires that (c=1 here)

\begin{equation}
E_{in}+M_{in}=\mathbf{p}_{in}^{2}/2M_{in}+M_{in}=E_{out}+M_{out}+E_{\gamma}=\left(\mathbf{p}_{in}-\mathbf{p}_{\gamma}\right)^{2}/2M_{out}+M_{out}+\left|\mathbf{p}_{\gamma}\right|
\end{equation}
where we have used the non-relativistic energy formula for the nuclei
energies. This can be rewritten as

\begin{equation}
\frac{\mathbf{p}_{\gamma}^{2}}{2M_{out}}+\left(1-\frac{\hat{\mathbf{k}}\cdot\mathbf{p}_{in}}{M_{out}}\right)\left|\mathbf{p}_{\gamma}\right|+\left(M_{out}-M_{in}\right)+\frac{\mathbf{p}_{in}^{2}}{2}\left(\frac{1}{M_{out}}-\frac{1}{M_{in}}\right)=0
\end{equation}
where I have used the substitution $\mathbf{p}_{\gamma}=\hat{\mathbf{k}}\left|\mathbf{p}_{\gamma}\right|$.
The last term in this equation is usually extremely small and can
be ignored. Solving for $\left|\mathbf{p}_{\gamma}\right|$ then yields

\begin{equation}
\left|\mathbf{p}_{\gamma}\right|=\left|\frac{-\left(1-\frac{\hat{\mathbf{k}}\cdot\mathbf{p}_{in}}{M_{out}}\right)+\sqrt{\left(1-\frac{\hat{\mathbf{k}}\cdot\mathbf{p}_{in}}{M_{out}}\right)^{2}-2\left(M_{out}-M_{in}\right)/M_{out}}}{1/M_{out}}\right|
\end{equation}

In the case where $\mathbf{p}_{in}=0$ we get

\begin{equation}
\left|\mathbf{p}_{\gamma}\right|=\left|-M_{out}+M_{out}\sqrt{1-2\left(M_{out}-M_{in}\right)/M_{out}}\right|
\end{equation}

which, on expanding the square root in a Taylor series, yields

\begin{equation}
\left|\mathbf{p}_{\gamma}\right|=M_{in}-M_{out}-E_{\mathcal{R}}
\end{equation}
where the recoil energy is $E_{\mathcal{R}}=\left(M_{in}-M_{out}\right)^{2}/2M_{out}$.
Because the mass of the nucleus changes in the decay, the kinetic
energy term changes to reflect this, and so does the Hamiltonian operator.
So, in a strict sense, the $\left|\mathbf{p}_{in}\right\rangle $
and $\left|\mathbf{p}_{out}\right\rangle $ are states in two different
Hilbert spaces such that for free particles we have

\begin{equation}
\frac{\mathbf{p}_{in}^{2}}{2M_{in}}\left|\mathbf{p}_{in}\right\rangle =E_{in}\left|\mathbf{p}_{in}\right\rangle 
\end{equation}

\begin{equation}
\frac{\mathbf{p}_{out}^{2}}{2M_{out}}\left|\mathbf{p}_{out}\right\rangle =E_{out}\left|\mathbf{p}_{out}\right\rangle 
\end{equation}
So, we must take the Hilbert space direct sum to construct one for
the entire system. In systems where the Mössbauer effect is observed,
the recoil energy is considerably larger than the decay width of the
nuclear decay. Thus a gamma ray emitted from a free nucleus cannot
be resonantly re-absorbed by a similar free nucleus in its ground
state. But Mössbauer noticed that the measured resonant absorption
in some circumstances where both source and absorber were in a solid
crystal was in contradiction with the free particle model of nuclear
decay, and in particular increased with temperature rather than decreased
as one would expect in the free particle model, because the Doppler
broadening of the gamma rays should tend to increase the likelihood
that a gamma particle has the energy to be resonantly absorbed \cite{lipkin_quantum_2007}(see
chapter 2.2). The idea that explained this was that the recoil momentum
must be taken up by a crystal volume containing many atoms surrounding
the decaying nucleus, and with a significant probability that no energy
is lost to the lattice due to creation of phonons in this process. 

For a bound nucleus decaying in a crystal, the dynamics of the crystal
come into play. The mass change in the decaying nucleus is typically
ignored because

\begin{equation}
M_{in}-M_{out}\ll M_{in}
\end{equation}
The standard Lippmann-Schwinger scattering theory is typically used
\cite{lippmann_variational_1950}, where the S matrix satisfies the
equation

\begin{equation}
S=U(\infty,-\infty)=1-\left(i/\hbar\right)\int_{-\infty}^{\infty}H_{1}(t)U(t,-\infty)dt
\end{equation}
where $H_{1}$ is the interaction Hamiltonian, and $U$ the time evolution
operator in the interaction picture. When considering the decay of
a radioactive particle, it is more appropriate to consider an approximate
S operator for a finite time interval T which is small compared to
the lifetime of the particle ($9.8\times10^{-8}$sec for the 14.4
keV excitation of iron-57 for example) but nevertheless assumed large
compared to the relaxation time of the crystal holding the excited
nucleus. 

\begin{equation}
S(T)=U(T/2,-T/2)
\end{equation}
Considering just the crystal state vector, one can expand the initial
state in terms of a complete set of momentum eigenstates for the decaying
nucleus, denoted by index L. The Hilbert space of the whole crystal
is a product space spanned by products of eigenstates for each particle
in the solid. We concentrate on the momentum of the decaying nucleus
and write

\begin{equation}
\left|in\right\rangle =\sum_{p_{L}'}\left|\mathbf{p_{L}}'\right\rangle \left\langle \left.\mathbf{p}_{L}'\right|in\right\rangle 
\end{equation}
The out state is obtained by acting with the S matrix operator

\begin{equation}
\left|out\right\rangle =S\left|in\right\rangle =\sum_{p_{L}',p_{L}''}\left|\mathbf{p_{L}}''\right\rangle \left\langle \mathbf{p_{L}}''\right|S\left|\mathbf{p_{L}}'\right\rangle \left\langle \left.\mathbf{p_{L}}'\right|in\right\rangle 
\end{equation}
The $\left|in\right\rangle $ state denotes the state of the crystal
at $t=-\infty$, and the $\left|out\right\rangle $ state denotes
the state of the crystal at $t=+\infty$. This S matrix formalism
does not address the issue of how long it takes for the solid to relax
into the final state after the decay photon is detected. If we concern
ourselves with the case that a single decay occurs, then momentum
conservation requires 

\begin{equation}
\mathbf{p}_{L}''=\mathbf{p}_{L}'-\mathbf{p}_{\gamma}
\end{equation}
and so

\begin{equation}
\left|out\right\rangle =\sum_{p_{L}'}\left|\mathbf{p_{L}}'-\mathbf{p}_{\gamma}\right\rangle \left\langle \mathbf{p_{L}}'-\mathbf{p}_{\gamma}\right|S\left|\mathbf{p_{L}}'\right\rangle \left\langle \left.\mathbf{p_{L}}'\right|in\right\rangle 
\end{equation}
but we can substitute the replacement $\left|\mathbf{p_{L}}'-\mathbf{p}_{\gamma}\right\rangle =exp(-i\mathbf{p}_{\gamma}\cdot\mathbf{x}_{L})\left|\mathbf{p_{L}}'\right\rangle $
so that

\begin{equation}
\left|out\right\rangle =\sum_{p_{L}'}\left|\mathbf{p_{L}}'-\mathbf{p}_{\gamma}\right\rangle =\sum_{p_{L}'}exp(-i\mathbf{\hat{p}}_{\gamma}\cdot\mathbf{\hat{x}}_{L})\left|\mathbf{p_{L}}'\right\rangle \left\langle \mathbf{p_{L}}'-\mathbf{p}_{\gamma}\right|S\left|\mathbf{p_{L}}'\right\rangle \left\langle \left.\mathbf{p_{L}}'\right|in\right\rangle 
\end{equation}
Now if we assume that the initial state is such that $\left\Vert \mathbf{p_{L}}'\right\Vert \ll\left\Vert \mathbf{p_{\gamma}}\right\Vert $,
so that $\left\langle \mathbf{p_{L}}'-\mathbf{p}_{\gamma}\right|S\left|\mathbf{p_{L}}'\right\rangle \approx\Bigl\langle-\mathbf{p}_{\gamma}\Bigr|S\Bigl|0\Bigr\rangle$,
then we have

\begin{equation}
\left|out\right\rangle =\Bigl\langle-\mathbf{p}_{\gamma}\Bigr|S\Bigl|0\Bigr\rangle exp(-i\mathbf{\hat{p}}_{\gamma}\cdot\mathbf{\hat{x}}_{L})\sum_{p_{L}'}\left|\mathbf{p_{L}}'\right\rangle \left\langle \left.\mathbf{p_{L}}'\right|in\right\rangle =\Bigl\langle-\mathbf{p}_{\gamma}\Bigr|S\Bigl|0\Bigr\rangle exp(-i\mathbf{\hat{p}}_{\gamma}\cdot\mathbf{\hat{x}}_{L})\biggl|in\biggr\rangle
\end{equation}
The main effect on the crystal is an increase in the momentum due
to the recoil of the decayed nucleus (see \cite{lipkin_quantum_2007}
eqn 3.1 and following for a more complete discussion). Note that the
recoil momentum is applied initially to the decaying nucleus only,
but then over time it must be spread around the rest of the crystal.

\section{A non-perturbative approach based on super-radiance theory in QED}

In an effort to clarify the physical mechanisms which enable the Mössbauer
effect to occur, Preparata together with others applied his ``super-radiance''
theory of coherent oscillation to the problem of the rigid recoil
\cite{bressani_what_1992,preparata_qed_1995}. They argued first off
that a better explanation is required for the Mössbauer effect than
the one physicists have developed. In fact the field was given over
to chemists long ago, and not much attention has been paid to it by
physicists for some time. They argued that in conventional solid state
physics models, at very short distances, the presence of the solid
can not have very much effect on the Mössbauer decay. They used the
term ``asymptotic freedom'' for this expected decoupling, although
this term is a bit confusing in this context since it has little to
do with the asymptotic freedom of non-abelian guage theories. What
they mean by ``asymptotic freedom'' is that when a system dominated
by longer range forces is perturbed by a short range disturbance,
it behaves as if it were free. Think of a bullet fired into a pendulum
in equilibrium. At the instant of impact, the harmonic restoring force
on the pendulum plays no role. He argues that the Mössbauer effect
is just such a case, and that we should expect the recoil of the emitted
(or absorbed) x-ray should be taken up entirely by the emitting nucleus
if the conventional picture of a solid is correct. The rigid stiffness
model of a volume of nuclei employed to explain the observed effect
is clearly in contradiction to this intuitively expected asymptotic
freedom. So they argued that conventional solid state physics really
suggests that the Mössbauer effect is simply impossible, and something
fundamentally different must be employed. They accuse mainstream physics
of ignoring this logical conundrum by accepting the rigid theory,
which does explain the experimental data, without questioning its
logical consistency. Personally I think they were correct in this
assessment. They argue that what is different is that the nuclei in
a solid can undergo a collective and coherent low amplitude collective
oscillation about their respective mean lattice sites, and that this
collective motion gives a small volume of a solid the rigidity that
is required to explain the recoil phenomenon. So Preparata et al.
treat the set of nuclei in a metal as if it were a plasma, and they
consider it quantum mechanically by introducing a wave-field for the
nuclei $\Psi(\overrightarrow{x},\overrightarrow{\xi})$, where $\overrightarrow{x}$
denotes the mean position of a nucleus, and $\overrightarrow{\xi}$
its deviation from this mean. These nuclei can oscillate collectively
at the plasma frequency

\begin{equation}
w_{p}=\frac{Ze}{\sqrt{M_{N}}}\sqrt{n}
\end{equation}

\noindent where $Z,\:M_{N},\:and\:n$ are the atomic number, nuclear
mass, and number density for the nuclei in the solid. This oscillation
is not a phonon effect, but is a plasmon effect, and Preparata et
al. argue that the quantization of this collective motion can lead
to the rigidity observed in the Mössbauer phenomenon. They do not
specifically calculate the time that it takes for the recoil to occur.
As the region of the solid is perfectly rigid though, one might expect
that this picture leads to an intantaneous recoil of a collective
group of nuclei.

\section{{\normalsize{}A standard model picture based on perturbation theory
and Feynman diagrams}}

The standard model of particle physics is considered at this time
to be the best and only all-encompassing fundamental description of
nuclear physics, atomic physics, and condensed matter physics known
\cite{oerter_theory_2006}. In condensed matter physics, it is for
the most part equivalent to quantum electrodynamics because the electromagnetic
interaction (via virtual photons) dominates the interactions. Since
the primary tool for solving problems in QED is perturbation theory
and Feynman diagrams, let's assume to start off that such a description
is suitable in this case, except that the nuclear decay dynamics themselves
involve the weak and strong forces too. Let us therefore assume first
that a perturbative approach, based on Feynman diagrams, can be used
to describe the Mössbauer process. In any nuclear decay of an isomer
which emits a photon, including a Mössbauer decay, the basic Feynman
diagram for the decay is shown in \prettyref{fig:The-basic-Feynman-diagram}.
The effects of the weak and strong forces are summarized in vertex
function for this diagram. 

\begin{center}
\begin{figure}[H]
\begin{centering}
\includegraphics[scale=0.5]{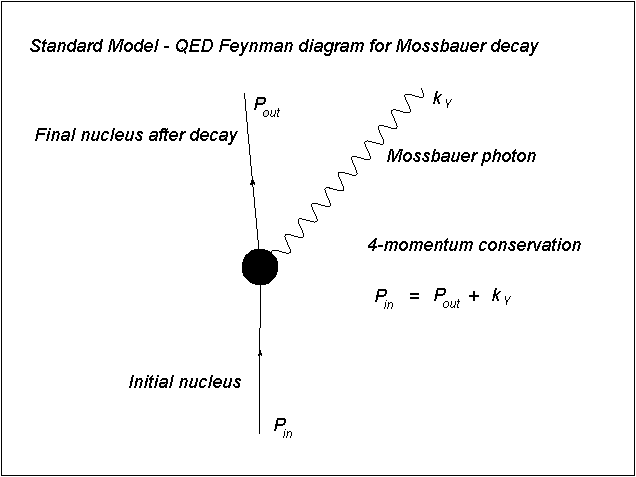}
\par\end{centering}

\protect\caption{A zoom-in on the core Feynman diagram for the Mössbauer process\label{fig:The-basic-Feynman-diagram}}
\end{figure}

\par\end{center}

In this elementary depiction of any nuclear decay by single photon
emission, the 4-momentum is conserved, and the three legs of the vertex
can in general be off the mass shell.

\begin{equation}
P_{in}^{\mu}=P_{out}^{\mu}+k_{\gamma}^{\mu}
\end{equation}

\begin{figure}[H]
\begin{centering}
\includegraphics[scale=0.5]{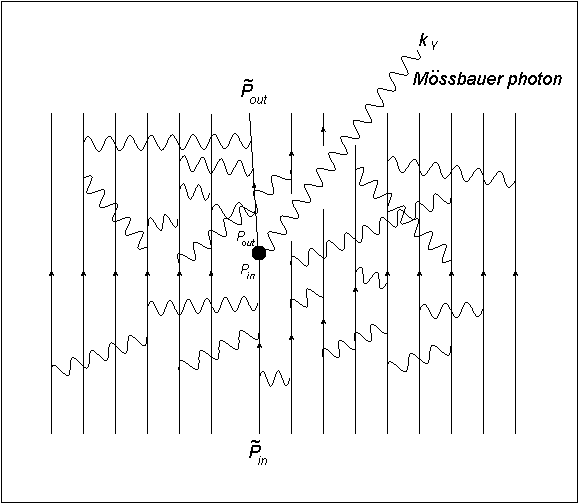}
\par\end{centering}

\protect\caption{Feynman diagram for the Mössbauer process showing multiple photon
interactions that can diffuse the recoil momentum throughout the solid.
The solid vertical lines represent charged particles in the solid
including electrons and nuclei.\label{fig:A_full_Feynman_diagram}}

\end{figure}

The symbol $P_{in}^{\mu}$ is the 4-momentum of the initial nuclear
isomer which is decaying, and $P_{out}^{\mu}$ is the 4-momentum of
the final state. Now, if perturbation theory can be used, the ongoing
interaction cannot influence the conservation of 4-momentum at the
core decay vertex as shown in \prettyref{fig:The-basic-Feynman-diagram}.
In \prettyref{fig:A_full_Feynman_diagram} we show an example of some
of the other interactions which will occur between the decaying nucleus
and the rest of the solid. 

We must sum over all conceivable Feynman diagrams of this type to
calculate the probability amplitude for arriving at any given final
state. In these more complex diagrams, let's assume that the x-ray
photon with momentum $k_{\gamma}$ is on-shell ($k_{\gamma}^{\mu}k_{\gamma\mu}=0$)
since otherwise it could not travel very far out of the solid, and
that it passes through the solid without scattering. Now the incoming
and outgoing momenta for the decaying nucleus long before and long
after the decay event can be changed by the interaction with the solid,
and therefore 

\begin{equation}
\widetilde{P^{\mu}}_{out}\neq P_{out}^{\mu}
\end{equation}

\noindent \begin{flushleft}
and 
\par\end{flushleft}

\begin{equation}
\widetilde{P^{\mu}}_{in}\neq P_{in}^{\mu}
\end{equation}

It is plausible to assume, however, that at least $\widetilde{P^{\mu}}_{in}$
will be very close to the mass shell, and that the deviation in the
kinetic energy will be on the order of the expected thermal fluctuations
which are proportional to $kT$, where $K$ is Boltzman's constant
and $T$ is temperature of the solid in Kelvin degrees. Let us define
two rest masses for the two nuclear isomers. I use units such that
$c=\hbar=1$ and the metric signature is (+,-,-,-). Let us define
the following masses of the two nuclear states in a vacuum. These
are just the standard mass values you find in a nuclear database.

\begin{equation}
M_{in}=rest\:mass\:of\:the\:unstable\:isomer\:state\:of\:the\:nucleus
\end{equation}

\begin{equation}
M_{out}=rest\:mass\:of\:the\:stable\:ground\:state\:of\:the\:nucleus
\end{equation}

In order for the Mössbauer photon to be resonantly absorbed, its energy
must satisfy the following equality with high accuracy. 

\begin{equation}
k_{\gamma}^{0}=M_{in}-M_{out}
\end{equation}

Let us consider Feynman diagrams such that the photon has this value,
and working in the rest frame of the solid, let's suppose that to
a good approximation, especially at low temperatures, that we have
the incoming mass on the mass shell so that

\begin{equation}
P_{in}^{\mu}P_{in\mu}=\widetilde{P^{\mu}}_{in}\widetilde{P}_{in\mu}=M_{in}^{2}
\end{equation}

\noindent where I use the standard Einstein convention here, and repeated
Greek indices are to be summed over the values (0,1,2,3) where the
zero index represents the time dimension. Therefore, if we ignore
the thermal and zero-point agitation of the decaying nucleus in the
solid for the time being, we can write

\begin{equation}
P_{in}^{\mu}=(M_{in},0,0,0)
\end{equation}

\begin{equation}
k_{\gamma}^{\mu}=\left(\left(M_{in}-M_{out}\right),\:(M_{in}-M_{out})\hat{n}\right)
\end{equation}

\noindent where $\hat{n}$ denotes the unit 3-vector direction of
the Mössbauer photon. Now we can calculate $P_{out}^{\mu}$ from 4-momentum
conservation.

\begin{equation}
P_{out}^{\mu}=P_{in}^{\mu}-k_{\gamma}^{\mu}=\left(M_{out},-(M_{in}-M_{out})\hat{n}\right)
\end{equation}

\noindent Now we can calculate the mass of the nucleus that has resulted
from the decay and emission of a Mössbauer photon (we use lower case
$m_{out}$ for this mass because it turns out to be off shell and
therefore different from $M_{out}$)

\begin{equation}
m_{out}=\sqrt{P_{out}^{\mu}P_{out\mu}}=\sqrt{M_{out}^{2}-(M_{in}-M_{out})^{2}}<M_{out}
\end{equation}

For all examples of the Mössbauer effect the following is a good approximation

\begin{equation}
M_{out}\gg M_{in}-M_{out}
\end{equation}

\noindent and consequently

\begin{equation}
m_{out}=M_{out}\sqrt{1-\frac{(M_{in}-M_{out})^{2}}{\left(M_{out}\right)^{2}}}\approx M_{out}\left(1-\frac{(M_{in}-M_{out})^{2}}{2\left(M_{out}\right)^{2}}\right)=M_{out}-\frac{1}{2}\frac{(M_{in}-M_{out})^{2}}{M_{out}}
\end{equation}

\noindent and the amount that the nucleus if off-shell is 

\begin{equation}
\triangle m_{out}=m_{out}-M_{out}=-\frac{1}{2}\frac{(M_{in}-M_{out})^{2}}{M_{out}}=-\frac{1}{2}\frac{\overrightarrow{k_{\gamma}}{}^{2}}{M_{out}}\label{eq:Mass change formula}
\end{equation}

\noindent So the nuclear rest mass is off-shell in the negative direction.
We get the same formula for the absorption case. The mass of the final
state isomer is slightly below the mass shell for that isomer when
a Mössbauer photon is absorbed. For the case of $^{57}Fe$ we have
$\left(M_{in}-M_{out}\right)\thickapprox14.4$keV, and $\triangle m_{out}\thickapprox-0.002$eV.

This off mass shell behavior is impossible to describe correctly within
the framework of conventional solid state theory which is based almost
exclusively on non-relativistic quantum mechanics, and on-mass-shell
Schrödinger equations. It offers another explanation for the logical
basis of the Mössbauer effect, but unlike the super-radiance theory,
it uses a known property of conventional relativistic quantum field
theory, namely the off shell behavior of virtual particles in a Feynman
diagram. It's surprising that this fact has not been pointed out before
in the literature, which to my knowledge it hasn't. Perhaps it is
because the off mass shell behavior of virtual particles in Feynman
diagrams has been considered a mathematical enigma since the on-mass-shell
particle states are considered to be a complete set of states in the
conventional picture of a solid. 

At first, just after the decay, the recoil momentum is fully absorbed
by a single nucleus, but then a transitory period of diffusion must
occur. This momentum density that is concentrated at a single nucleus
must be spread throughout the entire lattice, and result in a state
corresponding to the common understanding of the Mössbauer effect
- that of a rigid crystal moving with a small recoil velocity. After
this diffusion occurs, the recoil momentum is described by this rigid
motion of the crystal, and presumably all of the nuclei return to
the mass shell (ie. their rest masses are the usual values). The excited
isomer will eventually decay back to the ground state too. The mechanism
for this diffusion presumably does not include any phonons since the
solid is excited with an energy that is below the threshold for producing
even one phonon. But it may include electromagnetic transport mediated
by response of the electron gas in the solid to the motion of the
recoil nucleus. 

I think that this view of the Mössbauer effect would be preferred
by particle physicists, but not by solid state physicists. Since there
are many more solid state physicists, it is probably the case that
the majority of physicists would reject the above description. In
my view it is needed to reconcile locality and causality with the
conventional interpretation of the Mössbauer effect. Since particle
physics is the parent field for nuclear physics, I think that the
standard model picture of the Mössbauer effect must be taken seriously.
Of course, nature may not work this way, and the more non-local interpretation
of solid state physicists could prevail if experiments could be devised
to rule out the above standard model picture. I would personally bet
on the standard model though. 

We must sum over all the possible final states of the solid, where
by final states I mean here states that have been reached some time
after the decay has occurred, so that the crystal has had time to
reach statistical equilibrium. I'm thinking in terms of just ordinary
QED here, so the Mossbauer photon can be taken on the mass shell (ie.
zero mass). Since the various configurations for the solid are not
measured, or even measurable, I would sum over amplitudes, following
Feynman's dictum for indistinguishable outcomes, and therefore interference
can occur. This is similar to the technique of summing over amplitudes
for emssion of infra-red photons in order to deal the inrared divergences
of QED. I think it will turn out if you do this that the final states
which are very near to the whole solid moving with fixed velocity
rigidly and with the proper recoil momentum will tend to be in phase
and constructively interfere if the energy of the crystal is below
the phonon gap for the crystal. Other final states would, I expect,
interfere with random phases, and so tend to cancel out. Thus the
Feynman paths in a path integral which resulted in the whole solid
moving rigidly with the required recoil momentum would be singled
out and have an enhanced amplitude, and this would enable one to calculate
the decay probability and width for them. This is how the phonon density
of states can enter the problem. 

Obviously, if this argument is correct, then transient off-mass-shell
behavior plays a critical role. Conventional solid state physics theory
has difficulty with describing this phenomenon. The only course of
action that I can see to remedy this situation is to modify solid
state physics to incorporate off-mass-shell states into the Hilbert
space of states. Such a possibility was long ago proposed by Greenberger
\cite{greenberger_theory_1970,greenberger_theory_1970-1,greenberger_useful_1974,greenberger_wavepackets_1974}.
One way to do this in a manifestly covariant way is to use the extensive
machinery of the Fock-Steuckelberg-Horwitz-Piron covariant quantum
mechanics \cite{fock_eigenzeit_1937,stueckelberg_remarque_1941,stueckelberg_signification_1941,horwitz_relativistic_1973,horwitz_relativistic_2015,fanchi_parametrized_1993}.
This theory is quite well developed, and seems well suited to study
this phenomenon. 

So in summary, the perturbative picture of the standard model says
the following about the Mössbauer effect:
\begin{enumerate}
\item All of the momentum of the emitted (absorbed) Mössbauer photon is
accounted for by the recoil of a single decayed (excited) nucleus
immediately after the decay (absorption) has occurred.
\item If the initial nuclear state is on its mass shell, then the final
nuclear state that results from and immediately after the Mössbauer
decay/absorption is off the mass shell.
\item The subsequent diffusion following a decay or absorption can in theory
propagate much faster than the speed of sound, up to the speed of
light, and result in a rigidly recoiled final state.
\end{enumerate}

\subsection{Digression into off-mass-shell covariant relativistic quantum mechanics}

The Feynman perturbation picture requires that either the final state
of the nucleus (or possibly even the initial state) must be off the
mass shell for a Mössbauer x-ray to be either emitted or absorbed
as in \ref{fig:The-basic-Feynman-diagram}. Thus, to handle this effect
rigorously, some sort of off-mass-shell version of quantum mechanics
must be used in which the on-mass-shell quantum states no longer form
a complete set of states. The most well-developed set of such theories
are of the Fock-Stueckelberg-Horwitz-Piron category. The recent very
fine book by Lawrence Horwitz \cite{horwitz_relativistic_2015} gives
a detailed account of most of the current status of this field. Sometimes
also referred to as the proper time formulation, the basic idea is
to add a second Lorentz invariant time variable. All particles then
move along trajectories in 4D Minkowski space which are parametrized
by a common universal time which is both a Lorentz invariant and a
common affine parameter for all trajectories, thus making the geometric
arena for physics 5 dimensional. A point particle in this space is
given a new name, and it's called an event. The new time plays much
the same role as Newtonian time in non-relativistic mechanics. The
theory was extended to interacting many-body systems in a landmark
paper by Horwitz and Piron in 1973 \cite{horwitz_relativistic_1973},
and this theory introduced new and interesting possibilities for quantum
mechanics. It's basically a Schrödinger type of theory, the equation
being first order in the new time but second order in the Minkowski
time variable. The particle coordinates, functions of the universal
time $\tau$, are 4 dimensional vectors. It led to a generalization
of electrodynamics called pre-Maxwell theory. The bibliography is
extensive, and can largely be found in \cite{horwitz_relativistic_2015}.
Methods for adding arbitrary spin have been worked out, and mechanisms
for the mass to return the the standard rest mass have been partially
explored. A recent paper by Martin Land \cite{land_speeds_2016} has
introduced a modification to pre-Maxwell theory that introduces a
new constant into the theory that controls how close it is to conventional
QED. These types of theories seem well suited for studying the relaxation
of a crystal back to equilibrium after a Mössbauer decay or absorption
has resulted in an off-mass-shell nucleus at a particular lattice
site in the solid. I won't attempt to give a thorough accounting of
this subject here, but simply include the basic Horwitz-Piron wave
equation for consideration:

\begin{equation}
\sum_{i=1}^{N}\left[\frac{\hbar^{2}}{2M_{i}}\left(p_{i}^{\mu}-q_{i}A^{\mu}\right)\left(p_{i,\mu}-q_{i}A_{\mu}\right)+V(x_{1},x_{2},..,x_{N})\right]\Psi(x_{1},...,x_{N},\tau)=i\hbar\frac{d\Psi(x_{1},...,x_{N},\tau)}{d\tau}
\end{equation}
where each $x_{i}$ is a 4-vector in Minkowski space, and $p_{i}^{\mu}=-i\hbar\partial/\partial x_{i,\mu}$,
$M_{i}$ the standard rest mass of each particle, $q_{i}$ their charge,
$A^{\mu}$ an external electromagnetic potential, and $V$ an arbitrary
potential function. In general in this theory, even in a region of
space time where $A^{\mu}=0$, and $V=0$, the particles can be off
the mass shell so that $p_{i}^{\mu}p_{i,\mu}\neq M_{i}^{2}$.

In a new paper \cite{land_speeds_2016}, Martin Land has developed
a modification the the pre-Maxwell theory which contains a new parameter,
he calls it $C_{5}$, which controls how closely the theory agrees
with standard QED, and also how quickly the masses of the particles
return to their standard rest mass values, once all interactions have
been removed. In a solid, one can imagine that Land's $C_{5}$ parameter
might depend on the materials properties of the solid, in the way
that the index of refraction does in conventional electromagnetic
theory. The response of the crystal to a Mössbauer decay or absorption
would then be controlled by this phenomenological parameter. This
off-mass-shell theory then would seem to be superior to the standard
Schrödinger equation for the many-body description of the solid in
this circumstance.

\section{{\normalsize{}Radiation from a charged metal sphere due to recoil
from the emission of a Mössbauer x-ray - a mechanism to measure response
time of Mössbauer recoil}}

Rather than try to settle the theoretical debate about the rigidity
of the Mössbauer crystal, I would like to propose an experiment for
consideration which might shed light on the subject. Consider a small
conducting sphere which consists of a material which is either undergoing
Mössbauer decays or is resonantly absorbing Mössbauer x-rays. Let
the sphere be charged with a total charge q. The charge will typically
reside on the surface, but this is not fundamentally necessary for
the effect. The effect that we will discuss is the same for either
sign of the charge, however a positivelly charged sphere may be more
stable to field emission than a negatively charged one, because of
the much higher probability of electron tunneling than ion tunneling
from a surface. If the entire sphere acts as a rigid body, as in the
usual (and perhaps naive) solid state physics picture of the Mössbauer
decay, then the conducting spherical shell of charge on the outer
surface of the sphere will experience a pulse of acceleration. Consequently
it will radiate electromagnetic energy away. The amount of radiated
energy will depend on how large the acceleration is. A stiff crystal
will lead to a short pulse of large acceleration, a diffusing crystal
will lead to a longer pulse of smaller acceleration. The stiffer the
acceleration, the more radiation is to be expected, because for the
lower frequencies ($\upsilon\ll R/c$) we expect Larmor's formula
to hold, and this states that the radiated energy is proportional
to the acceleration squared. The radiated power also varies as the
square of the charge, and so the signal can be enhanced by increasing
the charge up to the electrostatic breakdown limit. Whatever the amount
of energy lost to radiation is, this energy must be subtracted from
the energy available for either emission or absorption of of a Mössbauer
photon.

\begin{figure}[H]
\begin{centering}
\fbox{\begin{minipage}[t]{1\columnwidth}%
\begin{center}
\includegraphics[scale=0.7]{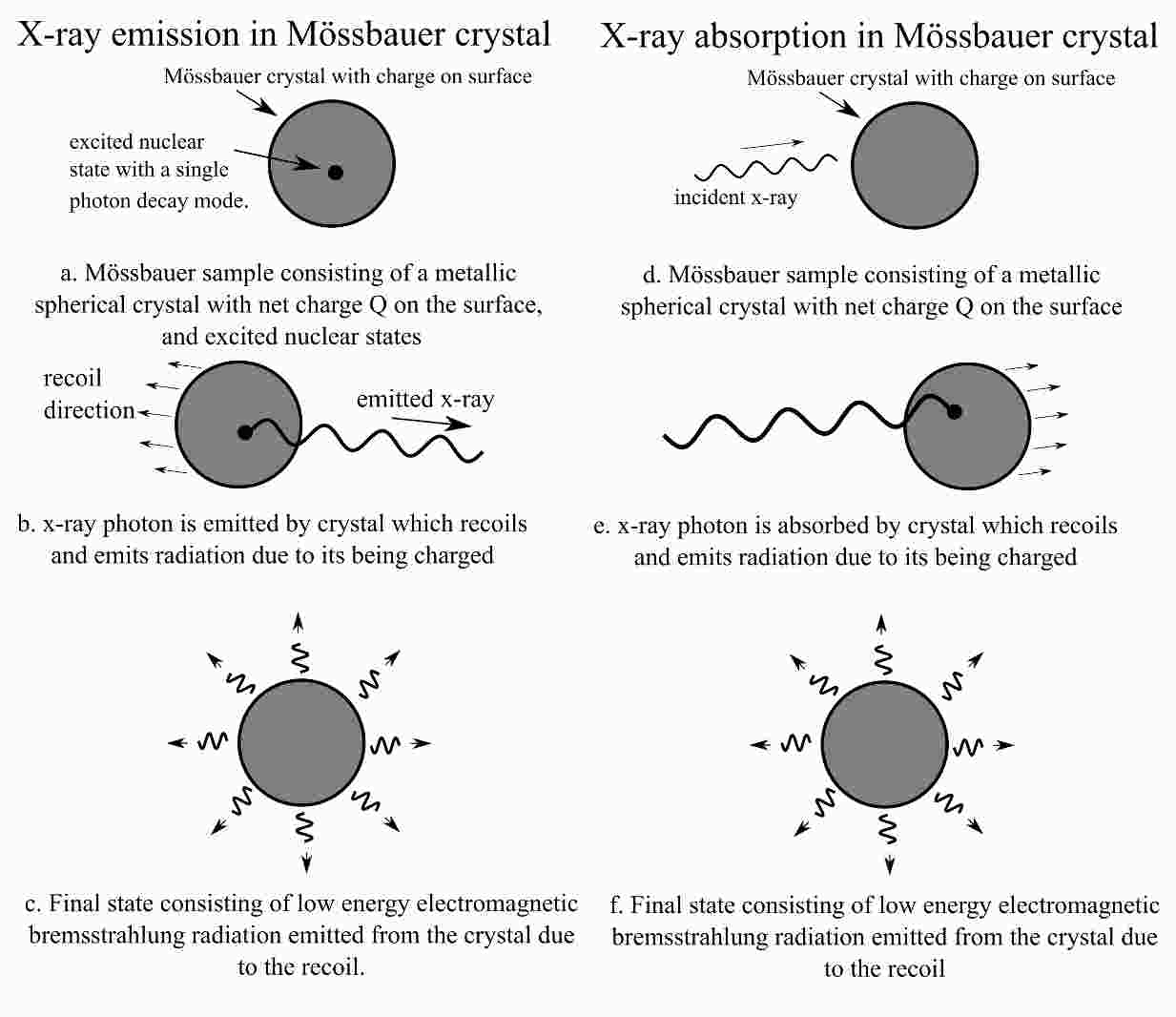}
\par\end{center}%
\end{minipage}}
\par\end{centering}

\noindent \begin{centering}
\protect\caption{Radiation emitted or absorbed by a charged sphere containing a decaying
nucleus embedded in a metal crystal }

\par\end{centering}

\end{figure}
Let's treat the recoiling crystal classically as a first approximation.
Let its acceleration along the recoil direction be be a(t). Then the
change in its velocity is 

\begin{equation}
v_{f}-v_{i}=\int_{-\infty}^{+\infty}a(t)dt
\end{equation}
If the direction and energy of the emitted or absorbed x-ray is known,
then the change in the velocity of the recoiling crystal is determined
by momentum conservation. As a first approximation to the energy radiated,
we can use Larmor's formula (in Gaussian cgs units)

\begin{equation}
P_{Larmor}(t)=\frac{2}{3}q^{2}a(t)^{2}/c^{3}\label{eq: Larmor's formula}
\end{equation}

We see that in the limit of an infinite acceleration, we must have
a(t) tending to a delta function in time, and so the radiated energy
given by 

\begin{equation}
E_{rad}=\int_{-\infty}^{+\infty}P_{Larmor}(t)dt=\frac{2}{3c^{3}}q^{2}\int_{-\infty}^{+\infty}a(t)^{2}dt
\end{equation}
will become infinite. So we must limit the recoil time to a nonzero
value. For simplicity, let's assume that the acceleration is constant
for a time T, and zero before and after this time, so that the aceleration
is a simple step function.

\begin{equation}
\mathbf{a}(t)=\mathbf{a}_{0},\:-T/2<t<T/2,\:and\:0\:otherwise
\end{equation}

or, in terms of the unit pulse function $\Pi$,

\begin{equation}
\mathbf{a}(t)=\mathbf{a}_{0}\Pi\left(\frac{t}{T/2}\right)\label{eq:Step Function acceleration}
\end{equation}

We must require that the integral of the acceleration pulse over all
time gives the change in the velocity of the crystal caused by the
recoil of the Mössbauer decay (here $M_{\chi}$ is the mass of the
stiff crystal, and $k_{\gamma}$ is the wave number of the x-ray,
either emitted or absorbed). Assuming that the initial momentum of
the crystal is zero, the magnitude of its final recoil momentum is
$\pm\hbar k_{\gamma}$, an so its final velocity after recoil is given
by

\begin{equation}
V_{M_{x}}=a_{0}T=\pm\hbar k_{\gamma}/M_{x}
\end{equation}

\noindent where the $+$ sign if for absorption of a photon, and the
$-$ sign is for emission. Integrating over time we find the total
energy radiated

\begin{equation}
E_{rad}=E_{Larmor}=\int_{-T/2}^{T/2}P(t)dt=\frac{2}{3}q^{2}Ta_{0}^{2}/c^{3}=\frac{2q^{2}}{3c^{3}}\frac{V_{M_{x}}^{2}}{T}=\frac{2q^{2}\hbar^{2}k_{\gamma}^{2}/M_{x}^{2}}{3c^{3}T}=\frac{2\hbar^{2}k_{\gamma}^{2}}{3c^{3}T}\left(\frac{q}{M_{\chi}}\right)^{2}\label{eq:Larmor pulse radiation}
\end{equation}

This energy must come from somewhere, implying that the energy of
the Mössbauer event must be reduced by this amount. \textcolor{black}{The
linewidth of the Mössbauer decay in iron-57 is about $1.0\times10^{-8}$eV,
and this determines the sensitivity of energy measurements. If the
radiative loss $E_{rad}$ is comparable or larger than this linewidth,
then the Mössbauer apparatus can in principle measure it. Let us calculate
the energy radiated if $T$ is the time it takes light to travel the
diameter of the particle. }

\subsection{Calculation of the energy radiated for speed of light recoil response
using iron-57 as the Mössbauer-active material with $T=2R/c$}

\textcolor{black}{To be specific, suppose the crystal consists of
iron-57, the most common crystal used in the Mössbauer effect. We
approximate the mass density iron-57 at room temperature by the corresponding
value of natural iron, including all stable isotopes, which is about
$7.2gm/cm^{3}$, and $k_{\gamma}=14.4$keV for $^{57}Fe$. The linewidth
of the Mössbauer decay in iron-57 is about}

\begin{equation}
\delta E=1.0\times10^{-8}eV
\end{equation}

\noindent \textcolor{black}{and this determines the sensitivity of
energy measurements. If the radiative loss $E_{rad}$ is comparable
or larger than this linewidth, then the Mössbauer apparatus can in
principle measure it. If $R$ is the radius of the sphere, we assume
here that $T=2R/c$, the time that it takes light to move the diameter
of the crystal.}

\noindent \textcolor{black}{Then we obtain}

\textcolor{black}{
\begin{equation}
E_{rad}=\frac{2}{3}\left(\frac{\hbar k_{\gamma}}{cM_{x}}\right)^{2}\left.\frac{q^{2}}{cT}\right|_{T=2R/c}=\frac{2}{3}\left(\frac{\hbar k_{\gamma}}{cM_{x}}\right)^{2}\frac{q^{2}}{2R}
\end{equation}
}but, the electrostatic energy of the charged sphere is just (in electrostatic
cgs units)

\begin{equation}
E_{es}=\frac{q^{2}}{2R}
\end{equation}
and therefore

\textcolor{black}{
\begin{equation}
E_{rad}=\frac{2}{3}\left(\frac{\hbar k_{\gamma}}{cM_{x}}\right)^{2}E_{es}
\end{equation}
}Now the kinetic energy of the recoiling crystal must be less than
$\delta E$ in order for the \textcolor{black}{Mössbauer effect to
work. So, assuming this, we can write}

\begin{equation}
\left(\hbar k_{\gamma}\right)^{2}/2M_{x}<\delta E\label{eq:Crystal mass constraint}
\end{equation}

\begin{equation}
E_{rad}<\frac{4}{3}\delta E\frac{E_{es}}{M_{x}c^{2}}\label{eq:Energy radiated at maximum}
\end{equation}
We can't make $R$ arbitrarily small, because we must statisfy the
constraint \prettyref{eq:Crystal mass constraint}. The crystal mass
must therefore satisfy the condition:

\begin{equation}
M_{x}>\left(\hbar k_{\gamma}\right)^{2}/2\delta E
\end{equation}
For Iron-57 we find

\begin{equation}
M_{x}>\left(14.4keV\right)^{2}/\left(2\times10^{-8}eV\right)=1.0368\times10^{16}eV=1.113\times10^{7}AMU
\end{equation}
Each iron atom has a mass of about $56.9$amu, so the crystal must
contain at least $1.9\times10^{5}$ iron atoms for the Mössbauer effect
to work. We can calculate the radius from the formula

\begin{equation}
M_{x}=\frac{4}{3}\pi R^{3}\left(7.2gm/cm^{3}\right)
\end{equation}

\begin{equation}
R_{max}\approx84.9\textrm{\AA}
\end{equation}
The question then is, what is the maximum amount of (positive) charge
that can be placed on an iron sphere of radius R before spontaneous
emission of iron ions occurs? In order to estimate this, we make some
crude assumptions. The threshold energy for sputtering of most metals
is usually in the range of 10 to 30 eV. Let $E_{th}$ be this energy
for iron-57. We assume that the maximum electric field strength in
the radial direction at the surface that can be withstood without
iron-57 ions being spontaneously emitted is approximately determined
by the condition

\begin{equation}
2e\left|\mathbf{E}_{max}\right|a_{x}=E_{th}
\end{equation}
where $a_{x}$ is the lattice spacing of the crystal ($\sim2.87\textrm{\AA}for\:iron$).
We give the iron ion a charge of +2 because each iron atom contributes
two electrons to the conduction band. So the maximum field strength
is, taking $10eV$ for the sputtering threshold:

\begin{equation}
\left|\mathbf{E}_{max}\right|=\frac{E_{th}}{2ea_{x}}
\end{equation}

\noindent Since we have spherical geometry, we have the relation just
outside the surface of the sphere

\begin{equation}
E_{es}=\frac{1}{2}QR\left|\mathbf{E}\right|
\end{equation}

\noindent the maximum value for $Q$ is determined by 

\begin{equation}
\left|\mathbf{E}_{max}\right|=\frac{Q_{max}}{4\pi\epsilon_{0}R_{max}^{2}}
\end{equation}

\noindent and the voltage of the charged sphere is

\begin{equation}
V_{mzx}=R_{max}E_{max}
\end{equation}

\noindent and therefore, the energy radiated is obtained from \prettyref{eq:Energy radiated at maximum}.
The results are:

\begin{equation}
R_{max}=85\mathring{A};\;E_{th}=10eV;\;\left|\mathbf{E_{max}}\right|=34.8GV/m;\;V_{max}=296V;\;Q_{max}=2.8\times10^{-16}C;\;E_{es}=258keV;\;M_{x}c^{2}=1.04\times10^{7}GeV
\end{equation}

\noindent and from these results we have the maximum radiation energy
loss 

\begin{equation}
E_{rad}\approx\frac{4}{3}\delta E\frac{E_{es}}{M_{x}c^{2}}=3.3\times10^{-19}eV
\end{equation}

\noindent and therefore

\begin{equation}
E_{rad}<<\delta E
\end{equation}
which implies that the effect is too small to measure. So the only
way that this effect could be measured is if the recoil happens essentially
instantaneously. Next we consider longer lifetime crystal materials.

\subsection{\label{sub:Calculation-for-Rhodium}Calculation of the energy radiated
for speed of light recoil response using Rhodium-103 as the Mössbauer-active
material with T=2R/c }

Rhodium-103 has a very long-lived isomer $^{103m}Rh$ which is a candidate
for precision Mössbauer research \cite{cheng_rhodium_2005,davydov_gravitational_2015}.
Its linewidth is $1.35\times10^{-19}eV$, which is eleven orders of
magnitude narrower than the $^{57}Fe$ isomer. The energy of the emitted
photon is $39.8$keV, and the mass of $^{103m}Rh$ is 102.9amu. The
density if $12.45gm/cm^{3}$. The lattice type is FCC, and the lattice
spacing is $3.8\textrm{\AA}$.The calculation is otherwise the same
as for $^{57}Fe$. In order for the Mössbauer effect to work, the
mass of the crystal in this case must be quite a bit larger then for
iron-57, because of the much narrower linewidth for Rhodium-103. This
larger mass then translates into a smaller radiation loss for a charged
crystal due to acceleration. Consequently, if we go through the calculation
done above for iron-57, but substituting the appropriate parameters
for Rhodium-103, we find that the radiated power is actually smaller,
so the reduced linewidth did not help. However, we shall now propose
a way to utilize the full advantage a longer lived Mössbauer element
such as Rhodium-103. We will consider an undersized Mössbauer crystal
for the absorber, so that the kinetic energy is too large to allow
for any Mössbauer absorption to occur. We propose to compensate for
this energy loss by a Doppler shift applied to the absorer or source
which would cancel the deficit in energy as far as the absorbed gamma
ray energy was concerned. The reduced crystal mass makes the acceleration
larger than it would be if the full Mössbauer crystal mass were used,
and thus the energy loss due to radiation and Larmor's formula will
be greater. Let us consider a crystal consisting of $N$ atoms of
a Mössbauer active material, and let $m$ denote the mass of a single
nucleus, and $q=N_{e}e$ .

\begin{equation}
M_{x}=Nm
\end{equation}

\noindent R is determined by the condition

\begin{equation}
Nm=\rho_{\mu}\frac{4}{3}\pi R^{3}\rightarrow R=\left(\frac{3mN}{4\pi\varrho_{\mu}}\right)^{1/3}
\end{equation}

\noindent where $\rho_{\mu}$ is the mass density. So we have

\begin{equation}
E_{rad}(N,N_{e})=\frac{2}{3}\left(\frac{\hbar k_{\gamma}}{cNm}\right)^{2}\frac{N_{e}^{2}e^{2}}{2}\left(\frac{4\pi\varrho_{\mu}}{3mN}\right)^{1/3}=\frac{N_{e}^{2}e^{2}}{N^{7/3}}\frac{1}{3}\left(\frac{\hbar k_{\gamma}}{cm}\right)^{2}\left(\frac{4\pi\varrho_{\mu}}{3m}\right)^{1/3}
\end{equation}

\noindent Plugging in numbers for rhodium-103 we obtain the following
formula

\begin{equation}
E_{rad}(N,N_{e})=\frac{N_{e}^{2}}{N^{7/3}}\left(5.5680\times10^{-13}eV\right)
\end{equation}

\noindent As an exmple, suppose that $N=10^{4}$, and $N_{e}=10^{3}$,
we calculate

\begin{equation}
E_{rad}(10000,1000)=2.584\times10^{-16}eV,\quad R=32\textrm{\AA},\quad V=450V,\quad v_{Doppler}=6.22mm/s
\end{equation}

\noindent where R is the radius of the sphere, V is the voltage at
the outside of the spherical nanoparticle, and $v_{Doppler}$ is the
velocity required so that the Doppler shift compensates for the energy
lost to kinetic energy of the recoil crystal, since we have intentionally
considered and udersized Mössbauer crystal. This energy shift is over
3 orders of magnitude larger than the decay linewidth, and so it should
be resolvable. Many other values of $N$ and $N_{e}$ give resolvable
radiation results. There are some possible problems with this estimate
though. One is that due to the small size of the crystal, the surface
effects might be important. On the surface there may be isomer shifts
on the energy levels of the rhodium-103 and rhodium-103m caused by
the modified lattice at the surface of the sphere. Also at the surface,
the outer several layers of atoms may experience a non-zero electric
field due to the net charge on the sphere. This field could cause
energy shifts in rhodium-103 and rhodium-103m because the electric
quadrupole moment for these nucleii may not be zero, and such quadrupole
moments are common among nucleii and cause energy shifts due to electric
fields \cite{stevens_nuclear_1976}. Other long-lived Mössbauer active
isotopes which might be considered as alternatives to rhodium are
$^{45}Sc$, $^{107}Ag$, and $^{109}A_{g}$ \cite{cheng_rhodium_2005}.
Although none of these have as narrow a linewidth, they are all more
studied in the literature.

In this section we assumed that $T=2R/c$, so that if the measured
value of $E_{rad}$ is greater than the estimated value here, the
diffusion rate is faster than the speed of light, and if it is lower,
then the diffusion rate is on average slower than the speed of light.

\subsection{Gedanken experiment for testing these ideas}

I am a poor experimentalist, but I can never forego the temptation
to design a thought experiment. Imagine a flat insulating disk, it
could be a plastic, a glass, or a metal on which is deposited on one
or both sides a layer of nano-particles of a controlled size of rhodium-103,
or some other Mössbauer active element. Place this disk inside of
a conducting tube which is held at some potential $V$ above ground.
The nano-particles are electromagnetically floating. Now shine a light
source whose wavelength is short enough to allow electrons to leave
the Mössbauer active nanoparticles and move to the concucting cylinder
by means of the photoelectric effect. Eventually, when equilibrium
is reached, the nano-particles will approach the voltage of the surrounding
tube, and will then have a net positive charge on them. Multiple disks
can be used to increase the absorption. Working in a non-ionizing
atmosphere like helium might be simpler than working in a vacuum,
provided the ionization levels could be kept to a low enough level.
See the accompanying figure. In order to have variable rates of Doppler
shift, I would propose a precision linear 1 axis motorized stage.
These are available with speeds up to about 300mm/s with high repeatability
and accuracy if cost is not an issue. They are routinely controlled
by software. By varying the voltage on the outer cylinder, the charge
on the nano-particles can be adjusted, and the energy shift caused
by charging the nano-particles can be measured by adjusting the velocity
of the stage to match the voltage on the cylinder.

\begin{figure}[H]
\begin{centering}
\includegraphics[scale=0.48]{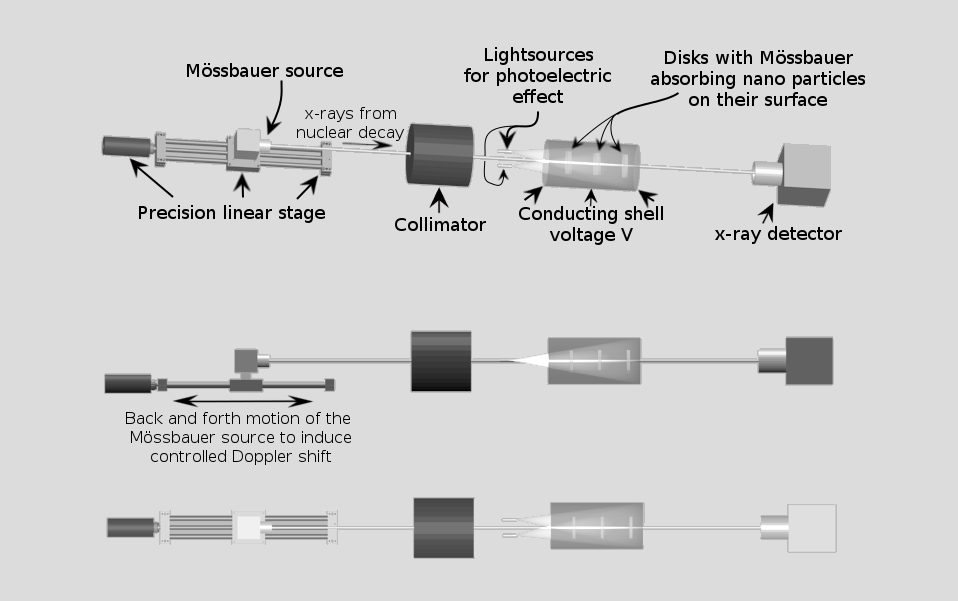}
\par\end{centering}

\protect\caption{Gedanken experiment for measuring the energy loss due to acceleration
of charged micro-spheres undergoing Mössbauer recoil, and thereby
gaining information about the rapidity of diffusion of momentum throughout
the crystal.}

\end{figure}

\section{Radiation from an arbitrary shaped rigid charged body in acceleration}

The Larmor formula is for a point charge. There are corrections to
it for an extended charge distribution which doesn't change shape
(ie. it moves rigidly) as it moves slowly along an arbitrary timelike
trajectory. To analyze this, let the charge density be

\begin{equation}
\rho(\mathbf{x},t)=\rho_{0}(\mathbf{x}-\mathbf{\boldsymbol{\xi}}(t))
\end{equation}

\begin{equation}
\int\rho(x,t)d^{3}x=1
\end{equation}

and where the current density is consequently

\begin{equation}
\mathbf{j}(\mathbf{x},t)=\mathbf{\boldsymbol{\dot{\xi}}}(t)\rho(\mathbf{x},t)
\end{equation}

The retarded electromagnetic potentials are (in vacuum, in the Lorentz
gauge, using Gaussian CGS units, and with c=1)

\begin{equation}
\varphi(\mathbf{x},t)=\int\frac{\rho(\mathbf{x}'-\boldsymbol{\xi}(t_{r}))}{\left|\mathbf{x}-\mathbf{x'}\right|}d^{3}x'
\end{equation}

\begin{equation}
A(\mathbf{x},t)=\int\frac{\mathbf{j}(\mathbf{x'},t_{r})}{\left|\mathbf{x}-\mathbf{x'}\right|}d^{3}x'
\end{equation}

where the retarded time is given by $t_{r}=t-\left|\mathbf{x}-\mathbf{x'}\right|/c$.
We assume that the charge distribution is localized so that for large
$R=\left|\mathbf{x}-\boldsymbol{\xi}(t-\left|\mathbf{x}\right|/c)\right|$
we have

\begin{equation}
\varphi(\mathbf{\mathbf{x}},t)=\frac{1}{R}\int\rho(\mathbf{\mathbf{x'}}-\boldsymbol{\mathbf{\xi}}(t_{r}))d^{3}x'+O(\frac{1}{R^{2}})=\frac{q}{R}+O(\frac{1}{R^{2}})
\end{equation}

\begin{equation}
A(\mathbf{x},t)=\frac{1}{R}\int\mathbf{\boldsymbol{\dot{\xi}}}(t_{r})\rho(\mathbf{x}'-\boldsymbol{\mathbf{\xi}}(t_{r}))d^{3}x'+O(\frac{1}{R^{2}})
\end{equation}

Note that $t_{r}$ depends on $\mathbf{x}^{'}$, and so in general
it cannot be taken outside the integral. The radiation $\mathbf{E}$
and $\mathbf{B}$ fields fall off as $1/R$ for large $R$. To calculate
them we first can calculate the $\mathbf{E}$ field, and then deduce
the $\mathbf{B}$ field from it. We see that $\nabla\varphi\propto1/R^{2}$,
and so it cannot contribute to the radiation field. Since $\mathbf{E}=-\mathbf{\nabla}\varphi-\partial\mathbf{A}/\partial t$,
we can approximate $\mathbf{E}=-\partial\mathbf{A}/\partial t+O(1/R^{2})$.

\begin{equation}
\mathbf{E}=-\partial/\partial t\left(\frac{1}{R}\int\mathbf{\boldsymbol{\dot{\xi}}}(t_{r})\rho(\mathbf{x}'-\boldsymbol{\mathbf{\xi}}(t_{r}))d^{3}x'\right)+O(1/R^{2})
\end{equation}

\begin{equation}
\mathbf{E}=-\frac{1}{R}\int\left[\mathbf{\ddot{\xi}}(t_{r})\rho(\mathbf{x}'-\boldsymbol{\mathbf{\xi}}(t_{r}))-\mathbf{\boldsymbol{\dot{\xi}}}(t_{r})\left(\mathbf{\boldsymbol{\dot{\xi}}}(t_{r})\cdot\mathbf{\nabla_{x'}}\rho(\mathbf{x}'-\boldsymbol{\mathbf{\xi}}(t_{r}))\right)\right]d^{3}x'+O(1/R^{2})
\end{equation}

Next we use the step function form for the acceleration which we assume
for the rigid Mössbauer acceleration \prettyref{eq:Step Function acceleration}.
We can take $\dot{\xi}$ to be a constant inside the integrand since
the recoil velocity of a Mössbauer crystal is very small. But in this
case the second term vanishes, and so we have simply

\begin{equation}
\mathbf{E}(\mathbf{x},t)=-\frac{\mathbf{a_{0}}}{R}\int\Pi(\frac{t_{r}}{T/2})\rho(\mathbf{x}'-\boldsymbol{\mathbf{\xi}}(t_{r}))d^{3}x'+O(1/R^{2})
\end{equation}

The magnetic field may be calculated from this by using $\mathbf{B}=\mathbf{\nabla}\times\mathbf{A}$
($\mathbf{\hat{R}}$ is a unit vector pointing from the charge to
the field point $\mathbf{x}$)

\begin{equation}
\mathbf{B}=\hat{\mathbf{R}}\times\mathbf{E}
\end{equation}

The Poynting vector in cgs units with c=1 is

\begin{equation}
\mathbf{S}=\frac{1}{4\pi}\mathbf{E}\times\mathbf{B}=\frac{1}{4\pi}\mathbf{E}\times\left(\mathbf{\hat{\mathbf{R}}\times\mathbf{E}}\right)=\frac{1}{4\pi}\left[\left(\mathbf{E\cdot E}\right)\hat{\mathbf{R}}-\left(\hat{\mathbf{R}}\cdot\mathbf{E}\right)\mathbf{E}\right]
\end{equation}

and the power radiated per solid angle is

\begin{equation}
d\mathcal{P}_{rad}=\mathbf{S\cdot dA}=\left(\hat{\mathbf{R}}\cdot\mathbf{S}\right)\mathbf{R}^{2}d\varOmega=\left[\mathbf{E}^{2}-\left(\hat{\mathbf{R}}\cdot\mathbf{E}\right)^{2}\right]\mathbf{R}^{2}\frac{d\varOmega}{4\pi}=\left[\mathbf{a_{0}}^{2}-\left(\hat{\mathbf{R}}\cdot\mathbf{a_{0}}\right)^{2}\right]\left(\int\Pi(\frac{t_{r}}{T/2})\rho(\mathbf{x}'-\boldsymbol{\mathbf{\xi}}(t_{r}))d^{3}x'\right)^{2}\frac{d\varOmega}{4\pi}+O(1/R^{2})
\end{equation}

Notice that this expression is never negative, so that the radiation
into any solid angle is greater than or equal to zero, as causality
requires. If we set the z axis to be parallel to $\mathbf{a_{0}}$,
we obtain

\begin{equation}
d\mathcal{P}_{rad}=\mathbf{a_{0}}^{2}sin^{2}(\theta)\left(\int\Pi(\frac{t_{r}}{T/2})\rho(\mathbf{x}'-\boldsymbol{\mathbf{\xi}}(t_{r}))d^{3}x'\right)^{2}\frac{d\varOmega}{4\pi}
\end{equation}

If we let D be the maximum extent of the charge distribution, and
then consider the limit cT>\textcompwordmark{}>D, we can replace $\varPi$
by 1, and the integral just yields the total charge so in this case

\begin{equation}
d\mathcal{P}_{rad}=\mathbf{a_{0}}^{2}sin^{2}(\theta)q^{2}\frac{d\varOmega}{4\pi}
\end{equation}

which on integration over the solid angle yields simply Larmor's formula
, with c=1. This result is independent of the shape of the charge
distribution. We see that in general the radiation rate is a fraction
of the Larmor rate, so that in general

\begin{equation}
d\mathcal{P}_{rad}=F(\Omega)\mathbf{a_{0}}^{2}sin^{2}(\theta)q^{2}\frac{d\varOmega}{4\pi}
\end{equation}

where

\begin{equation}
F(\Omega)=\frac{\left(\int\Pi(\frac{t_{r}}{T/2})\rho(\mathbf{x}'-\boldsymbol{\mathbf{\xi}}(t_{r}))d^{3}x'\right)^{2}}{q^{2}}
\end{equation}

if $\rho(x)$ has the same sign for all positions, then $F(\Omega)$
takes on values between 0 and 1. The total power radiated is then
a fraction of the Larmor total power

\begin{equation}
P_{rad}(t)=\int F(\Omega)\mathbf{a_{0}}^{2}sin^{2}(\theta)q^{2}\frac{d\varOmega}{4\pi}=fP_{Larmor}(t),\;0\leq f\leq1
\end{equation}

We note that rigid non-radiating accelerations are possible for certain
special classical systems as shown by multiple authors \cite{bohm_self-oscillations_1948,goedecke_classically_1964,schott_electromagnetic_1933,pearle_when_1978}.

We next consider the case of a spherical charged shell.

\subsection{Radiation from a rigid thin charged spherical shell}

For the case of a charged metal nanosphere, the charge distribution
is on the surface, and this can be approximated by a charged spherical
shell. The calculation of the radiation from a charged hollow spherical
shell has been studied extensively, as in \cite{bohm_self-oscillations_1948,yaghjian_relativistic_1992,yaghjian_relativistic_2005,rohrlich_dynamics_1997}
and in references therein. The shell moves rigidly, and the charge
density is radially symmetric in its rest frame. The vector coordinate
for its center is $\mathbf{\xi}(t)$. We need be concerned here with
only non-relativistic velocities. The charge density and current for
a general radial density, not necessarily a shell, are then given
by

\begin{equation}
\rho(\mathbf{x},t)=qf(\left|\mathbf{x}-\mathbf{\xi}(t)\right|),\quad\int d^{3}xf(\left|\mathbf{x}\right|)=1
\end{equation}
For a shell we have

\begin{equation}
f(\mathbf{x})=\frac{\delta(\left|\mathbf{x}\right|-R)}{4\pi R^{2}}
\end{equation}

\begin{equation}
j(x,t)=q\dot{\xi}(t)f(\left|\mathbf{x}-\mathbf{\xi}(t)\right|)
\end{equation}
An approximate formula for the electromagnetic self force on the shell
in the rest frame is given by Yaghjian \cite{yaghjian_relativistic_1992,yaghjian_relativistic_2005}(see
appendix A). His result is (in SI units)

\begin{equation}
\mathbf{F}_{self}(t)=\frac{-q^{2}}{6\pi\varepsilon_{0}Rc^{2}}\dot{\mathbf{u}}+\frac{q^{2}}{6\pi\epsilon_{0}c^{3}}\ddot{\mathbf{u}}+O(R),\;\mathbf{u}=0
\end{equation}
He also calculates the power done on the shell by this self force
to be

\begin{equation}
P_{el}(t)=\frac{-5q^{2}}{24\pi\epsilon_{0}Rc^{2}}\mathbf{u\cdot\dot{u}}+\frac{q^{2}}{6\pi\epsilon_{0}c^{3}}\mathbf{u\cdot\ddot{u}}+O(R),\:and\:for\:\frac{\mathbf{u^{2}}}{c^{2}}\ll1\label{eq: Yaghjian result}
\end{equation}
where $\mathbf{u}=d\mathbf{\mathbf{\xi}(t)}/dt$. Notice that $P_{el}(t)\neq u\cdot\mathbf{F}_{self}(t)$.
This discrepancy is explained carefully in \cite{yaghjian_relativistic_1992,yaghjian_relativistic_2005}.
One needs to include Poincare stresses to eliminate it. If one does
this, then the power equation is modified to (see \cite{yaghjian_relativistic_1992},
equation (5.5b), and taken to the small velocity limit)

\begin{equation}
P_{el}(t)=\frac{-q^{2}}{6\pi\epsilon_{0}Rc^{2}}\mathbf{u\cdot\dot{u}}+\frac{q^{2}}{6\pi\epsilon_{0}c^{3}}\mathbf{u\cdot\ddot{u}}+O(R),\:and\:for\:\frac{\mathbf{u^{2}}}{c^{2}}\ll1
\end{equation}
If we integrate the second term in this expression, and assume that
the acceleration vanishes in the distant future and distant past,
then we find

\begin{equation}
\int_{-\infty}^{\infty}\mathbf{u\cdot\ddot{u}}dt=-\int_{-\infty}^{\infty}\mathbf{\dot{u}^{2}}dt
\end{equation}
so the second term just gives us the Larmor formula result for the
energy radiated. The first term is a correction term which can be
interpreted as a mass renormalization. It can bbe rewritten as

\begin{equation}
\frac{-q^{2}}{6\pi\epsilon_{0}Rc^{2}}\mathbf{u\cdot\dot{u}}=\frac{-q^{2}}{6\pi\epsilon_{0}Rc^{2}}\frac{1}{2}\frac{d}{dt}\left(\mathbf{u}^{2}\right)=-\frac{d}{dt}\left(\frac{1}{2}\left(\frac{4}{3}m_{es}\right)\mathbf{u}^{2}\right)
\end{equation}
where

\begin{equation}
m_{es}=\frac{q^{2}}{8\pi\epsilon_{0}Rc^{2}}\textrm{ in SI units, and }m_{es}=\frac{q^{2}}{2Rc^{2}}\textrm{ in cgs Gaussian units}
\end{equation}

This correction term to the Larmor formula varies as the charge squared,
just as the radiation term does. This term cannot generally be ignored,
since it can be larger than the Mössbauer linewidth. But in principle,
the charge and the radius of the nanosphere can be known, and so it
is straightforward to calculated the magnitude of this term and take
its effect into account.

Of course the interior of the charged sphere is not empty in our case,
and depending on the materials involved, there would be additional
corrections which would require a numerical method to solve and are
beyond the scope of the current paper. Moreover, there are higher
terms in powers of $R$ that might play a role.

The methods of Bohm and Weinstein \cite{bohm_self-oscillations_1948}
allow one to estimate higher order (in powers of R) corrections to
the self force. Denote the $k^{th}$ Fourier component of a function
$g(x,t)$ as

\begin{equation}
g(k,t)=\frac{1}{\left(2\pi\right)^{3/2}}\int g(x,t)exp(-i\mathbf{k\cdot x})d^{3}x,\;k=\left|\mathbf{k}\right|
\end{equation}
and Defining $f(k,t)$ by

\begin{equation}
f(k,t)=\frac{1}{\left(2\pi\right)^{3/2}}\int f(\left|\mathbf{x}\right|,t)exp(-i\mathbf{k\cdot x})d^{3}x,\;k=\left|\mathbf{k}\right|
\end{equation}
The self force for small velocities but arbitrarily large acceleration
is found in \cite{bohm_self-oscillations_1948} (in Gaussian units)
to be

\begin{equation}
\mathbf{F}_{self}(t)=q^{2}\int_{0}^{\infty}d\tau G(\tau)\mathbf{\ddot{\boldsymbol{\xi}}}(t-\tau)
\end{equation}
where

\begin{equation}
G(\tau)=\frac{-32\pi}{3c}\int_{0}^{\infty}dkk\left|f(k)\right|^{2}sin(ck\tau)
\end{equation}
For a spherical shell of radius R (from equation (17) and following
in \cite{bohm_self-oscillations_1948})

\begin{equation}
f(\mathbf{x})=\frac{\delta(\left|\mathbf{x}\right|-R)}{4\pi R^{2}},\;f(k)=\frac{1}{\left(2\pi\right)^{3/2}}\frac{sin(kR)}{kR}
\end{equation}
and it follows that

\begin{equation}
G(\tau)=\left\{ \begin{array}{c}
-1/(3R^{2}c),\;\tau<2R/c\\
0,\;\tau>2R/c
\end{array}\right.
\end{equation}
Using \prettyref{eq:Step Function acceleration}

\begin{equation}
\mathbf{F}_{self}(t)=q^{2}G(0)\int_{0}^{2R/c}d\tau\mathbf{\ddot{\boldsymbol{\xi}}}(t-\tau)=q^{2}G(0)\int_{t-2R/c}^{t}dt'\mathbf{\ddot{\boldsymbol{\xi}}}(t')
\end{equation}

\begin{equation}
\mathbf{F}_{self}(t)=-q^{2}\left(\dot{\mathbf{\xi}}(t)-\dot{\mathbf{\xi}}(t-2R/c)\right)/(3R^{2}c)
\end{equation}
one can calculate higher order powers of $R$ corrections to the radiated
power. Once again, it is wrong in general to equate the radiated power
to the integrated work done by this self force.  The power equation
will have corrections as well, and these might have to be examined
if a serious program to study this effect experimentally, but we leave
this to future work.

\section{Conclusion}

The problem or at least the mystery of crystal rigidity in the Mössbauer
phenomenon has never been sufficiently clarified in the physics literature.
The subject touches on fundamental questions for relativity and quantum
mechanics. In this paper a partial sampling of the theoretical possibilities
was briefly reviewed. Part of the reason for the lack of resolution
has been the absence of a direct experiment that could measure the
intricacies of the diffusion of momentum inside the crystal immediately
after a Mössbauer event (either decay or absorption). To address this,
I have proposed here an experimental arrangement that utilizes the
extreme precision of the Mössbauer effect itself to approximately
measure the time duration of the impulse given to a small Mössbauer
crystal. This is achieved by measuring the amount of energy that is
radiated away by bremsstrahlung of a charged particle undergoing a
Mössbauer event. Besides helping to resolve the rigidity question,
this experimental technique might prove useful for measuring some
chemical or material properties, and thereby result in new and useful
analytical tools for chemistry and material science. The prediction
is that the radiated energy, and consequently the associated Mössbauer
energy shift, should vary in proportion to the charge squared divided
by the diffusion time as in \prettyref{eq:Larmor pulse radiation}.
If an experimental effect can be observed, then subsequent more detailed
theoretical and experimental analysis can zero in on the diffusion
dynamics of a Mössbauer event, and hopefully clarify the underlying
physical mechanism.

The treatment of bremsstrahlung radiation here is strictly classical.
There are undoubtedly quantum corrections to my analysis here, and
if experiments reveal that this effect is indeed measurable, then
it would justify plunging more deeply into a full quantum-mechanical
description.
\begin{acknowledgements}
I would like to acknowledge useful discussions and/or correspondence
with Lawrence Horwitz, Martin Land, Vladimir Kresin, and Robert Perlmutter.
\end{acknowledgements}

\bibliographystyle{spphys}
\bibliography{Mossbauer_effect,Mark_Davidson,Standard_model,Many_Body_Theory,Lippmann_Schwinger_Scattering,non_radiating_particles,Variable_mass_theory,LENR,Classical_Electron_Model,Martin_Land,Yaghjian}

\end{document}